\documentclass{article}
\usepackage[sort]{natbib}
\usepackage{amsmath}
\usepackage{amssymb}
\usepackage[normalem]{ulem}
\usepackage{tikz}
\usepackage{acronym}
\usetikzlibrary{arrows}

\usepackage{literate}
\usepackage{textboxes}
\usepackage{wrapfig}
\usepackage{subcaption}
\usepackage{hyperref}
\hypersetup{colorlinks=true,allcolors=uoeblue}
\newcommand{\term}[1]{\texttt{\color{uoeblue}\small #1}}
\usepackage{longtable}

\title{Compiling Combinatorial Genetic Circuits with Semantic Inference}
\author{William Waites, Goksel Misirli, Matteo Cavaliere,\\Vincent Danos, Anil Wipat}
\begin{document}
\maketitle
\begin{abstract}
  A central strategy of synthetic biology is to understand the basic processes
  of living creatures through engineering organisms using the same building
  blocks.
  Biological machines described in terms of parts can be studied by computer
  simulation in any of several languages or robotically assembled \emph{in
    vitro}.
  In this paper we present a language, the \acf{GCDL} and a compiler, the
  \acf{GCC}. This language describes
  genetic circuits at a level of granularity appropriate both for automated assembly
  in the laboratory and deriving simulation code.
  The \ac{GCDL} follows \acl{SW} practice and the compiler makes novel use of
  the logical inference facilities that are therefore available.
  We present the \ac{GCDL} and compiler structure as a study of a tool for
  generating $\kappa$-language simulations from semantic descriptions of
  genetic circuits.
\end{abstract}

\acrodef{GCDL}[GCDL]{Genetic Circuit Description Language}
\acrodef{GCC}[GCC]{Genetic Circuit Compiler}
\acrodef{RDF}[RDF]{Resource Description Framework}
\acrodef{SW}[SemWeb]{Semantic Web}
\acrodef{BNGL}[BNGL]{BioNetGen Language}
\acrodef{SBOL}[SBOL]{Synthetic Biology Open Language}
\acrodef{URI}[URI]{Universal Resource Identifier}
\acrodef{KBBF}[KBBF]{Kappa BioBricks Framework}
\acrodef{RBMO}[RBMO]{Rule-Based Model Ontology}
\acrodef{SKOS}[SKOS]{Simple Knowledge Organization System}
\acrodef{RDFS}[RDFS]{RDF Schema}
\section{Introduction}

Synthetic biology extends classical genetic engineering with concepts of
modularity, standardisation, and abstraction drawn largely from computer
engineering. The goal is ambitious: to design complex biological systems,
perhaps entire genomes, from first principles~\citep{baldwin2012synthetic}.
This enterprise has met with some success such as the microbial production of
drug synthesis~\citep{paddon2013,galanie2015}, new biofuels
production~\citep{ferry2012} and alternative approaches to disease
treatment~\citep{ruder2011}. However, most applications are still small and mostly
designed manually.

The are several obstacles to designing more complex circuits. The design
space of potential circuits is very large. Even when a design is chosen, there
is large \emph{a priori} uncertainty about what its behaviour will be. In many
cases the available information about molecular interactions in a cell is incomplete.
A secondary obstacle is that designs can be brittle and very sensitive to the host
environment in which they execute. In this context computational techniques become
important for identifying biologically feasible solutions to problems of biological system synthesis.

\begin{wrapfigure}{r}{0.6\textwidth}
  \includegraphics[width=0.55\textwidth]{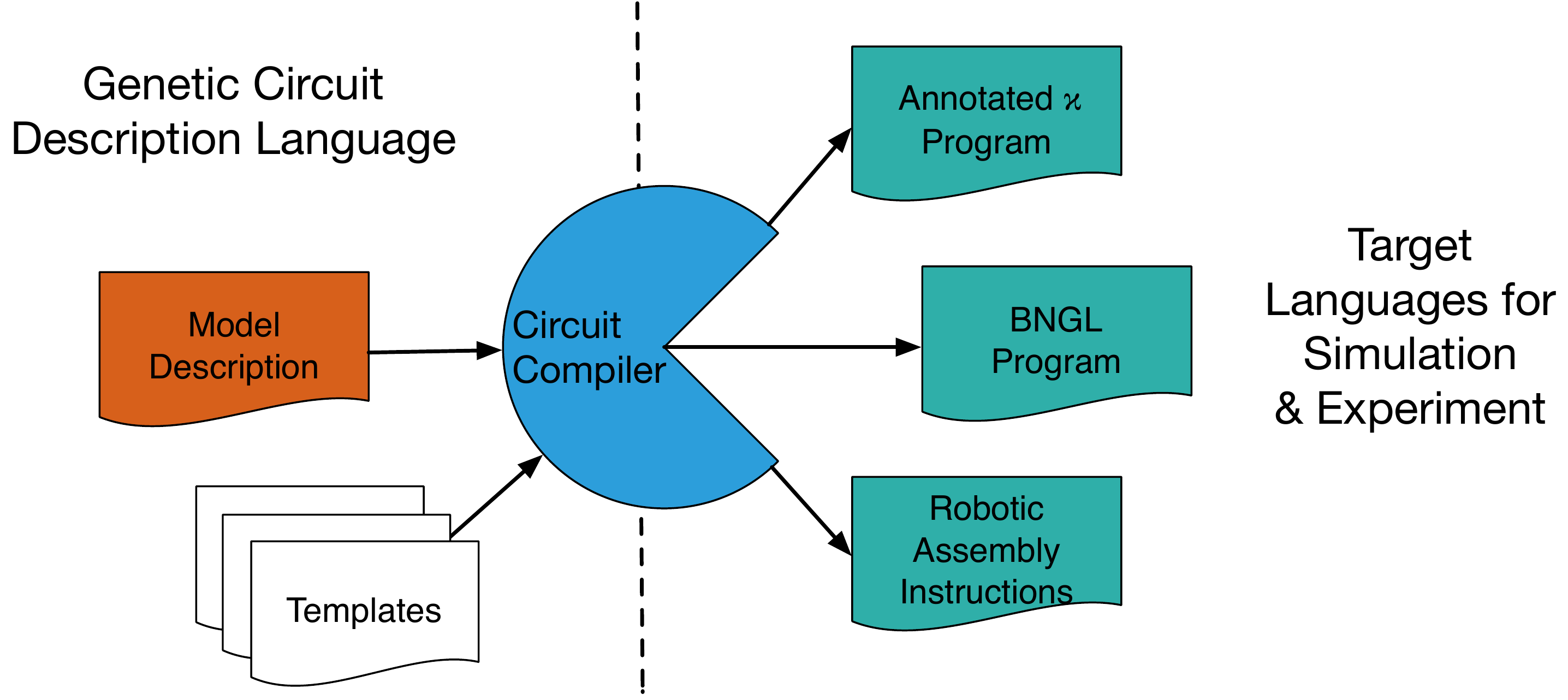}
  \caption{High-level data flow through the compiler. The compiler for synthetic
    gene circuits takes a model description written in \ac{GCDL}  and, using
    language-appropriate appropriate templates, creates code for simulation and
    laboratory assembly. We have implemented templates for annotated-$\kappa$ for
    the KaSim software, and envision similar for the \ac{BNGL} as well as
    \ac{SBOL}.}
    \label{fig:compiler-flow}
\end{wrapfigure}

In this paper we present a contribution to the computational infrastructure for synthetic biology, as a
language for describing genetic circuits, called the \acf{GCDL}, and a compiler
for translating them into programs, shown in
Figure~\ref{fig:compiler-flow}.
The \ac{GCDL} is an \acs{RDF}~\citep{cyganiak_rdf_2014} vocabulary which
facilitates gathering and collation of information about the constituent parts
of a genetic circuit~\citep{neal2014reappraisal}.
We use a strategy of contextual reasoning to obtain succinct input and flexible
output.
The output programs can be
specialised to various languages, such as the
KaSim flavour of $\kappa$~\citep{danos_rule-based_2007,krivine_kasim_2017},
BioNetGen's \ac{BNGL}~\citep{blinov2004bionetgen,harris_bionetgen_2016},
other representations such as \ac{SBOL}~\citep{galdzicki2014synthetic}
or indeed whichever form is required by robotic laboratory
equipment that assembles circuits \emph{in vitro}.
This output flexibility is accomplished using \emph{templates} that use facts
derived by inference rules~\citep{berners-lee_rdf_2005} from the input model;
input terms have meaning defined in terms of inference rules.

There are several reasons to automate the construction of 
simulation programs for complex genetic circuits over and
above the huge design space and associated uncertainties.
Writing these programs by hand is time-consuming and error prone, and there are
very few tools available for verification and debugging them. 
Descriptions of models in terms of simulation code are necessarily tightly
coupled to the interpreter of the simulation program's language.
This means that using a different interpreter or even different hardware is not
possible.

We solve  these problems by providing a high-level, modular,
implementation-independent language for describing gene circuits.
Code generation from this high-level description to a low-level language for
simulation greatly reduces the scope for error in coding simulations.
Because the language
is implementation-independent it is not tightly coupled to any particular
interpreter or hardware. In this way \ac{GCDL} facilitates
\emph{evergreen models}, models that are specified sufficiently well to be
unambiguous but not so specifically that they can only be executed or
constructed in one software package or environment.

\begin{wrapfigure}{r}{0.5\textwidth}
  \includegraphics[width=0.45\textwidth]{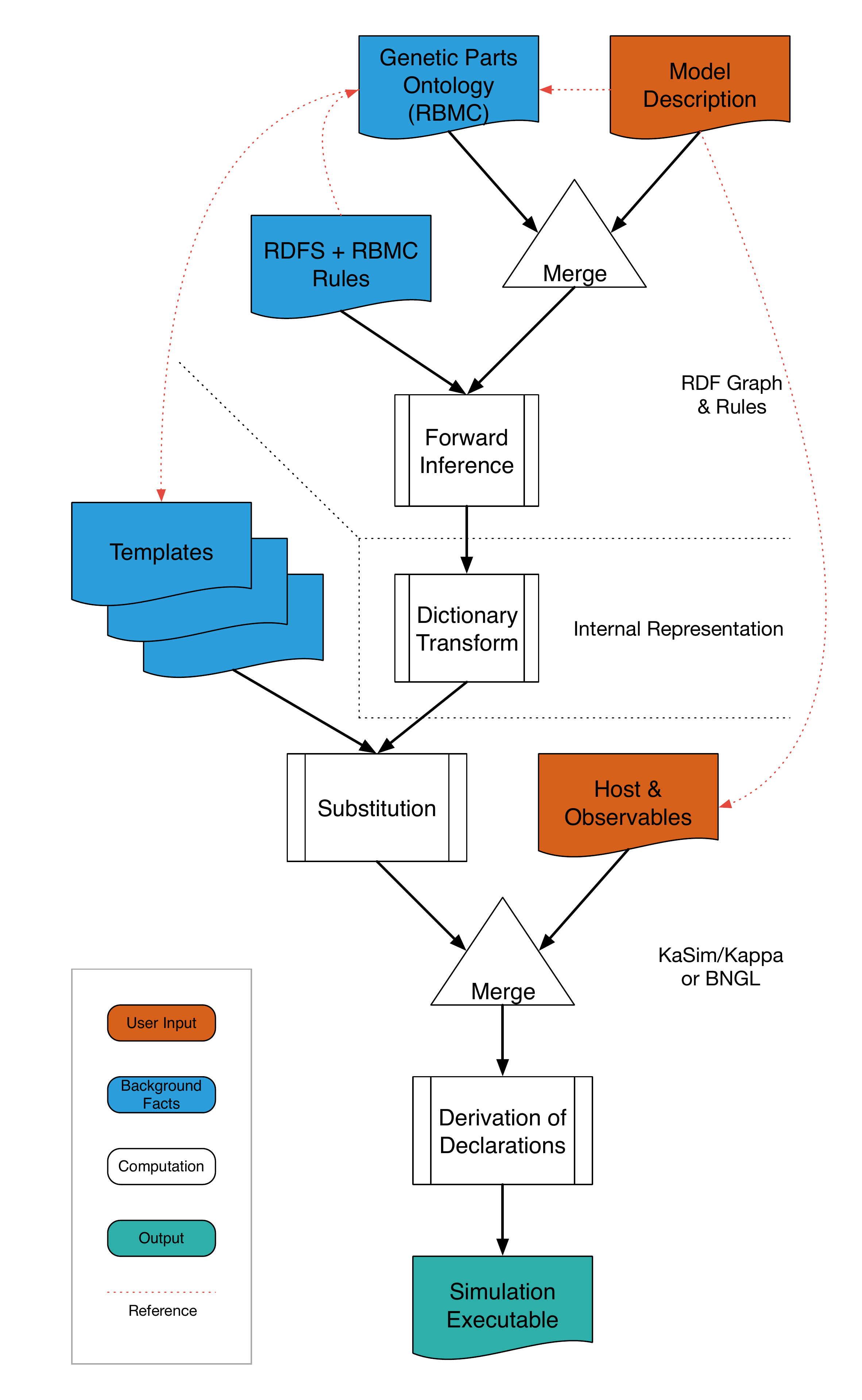}
  \caption{Detailed data flow through the compiler. This illustrates the use of
    inference to expand the \ac{GCDL} model to derive consequent information
    appropriate to producing the next stage of output in the
    specific target language.}
  \label{fig:compiler-flow-detail}
\end{wrapfigure}

The design of the compiler is general, and not limited to the present context of
genetic circuits. The design
shown schematically in 
Figure~\ref{fig:compiler-flow-detail}.
Domain specific languages and examples of compilers processing these languages
have previously been shown~\citep{pedersen_towards_2009,Beal2011,cai2011high,Hallinan2014}.
These languages are designed to allow for simulations using a particular
methodology such as solving systems of ordinary differential equations or using
Monte-Carlo simulation. 
Unlike previous approaches, we emphasise the use of abstraction to facilitate
\emph{retargeting} or production of output suitable for different simulation
environments and techniques as well as automated circuit assembly  in the
laboratory from a single description. Compiler
targets are implemented using conditional inference, essentially defining the
semantics of the terms used in the description of the circuit in a way that
is determined by the desired output type.

Even in a given target environment there are choices to be made about what level
of detail to treat a given phenomenon that depends on the question of biological
interest. Degradation of protein chains, for example, can be modelled as a
rule that simply destroys proteins at rate $k$, $P() \xrightarrow{k} \emptyset$. It can also
be modelled as a much larger set of rules to simulate the action of a protease
molecule attaching to the polymer and moving along it, disassembling each
constituent protein in turn.
If the  difference between these representations is large enough and relevant
enough to justify the extra computation involved with a larger rule-set, 
compiler output can be instructed to use different  \emph{templates}.
In this way simulations can be tuned to focus the questions of biological
interest, abstractions chosen such that more detailed modelling is done where
necessary and relevant, and those parts that are uninteresting can simply be
appropriately approximated.

We now proceed as follows.
In Section~\ref{sec:background} we give an overview
of those aspects of synthetic biology and genetic engineering that are necessary
to contextualise our work.
Next, in Section~\ref{sec:gcdl}, we explain the representation of this kind
of genetic circuit model in \ac{GCDL}, this is the main input to the compiler.
In order to understand the desired output of the
compiler, in Section~\ref{sec:ruleoutput} we show how these constructs
are represented as rule-based code for the $\kappa$ language simulator,
KaSim. There follows a
discussion in Section~\ref{sec:gcc} of how the compiler infers the executable
model from the input description. 
Finally, in Section~\ref{sec:ga} we sketch a
possible technique to answer the question of how to obtain a genetic circuit
from a given desired functional specification that uses these tools.

\section{Background}
\label{sec:background}

Let us begin from first principles. The Central Dogma of Molecular Biology has
it that genetic information flows, through chemical reactions, from nucleic
acids to proteins~\citep{crick1970central}. Under normal circumstances, such as
those that we are concerned with here, information is transferred from a DNA
sequence to a corresponding RNA sequence by the action of an RNA polymerase
molecule, and thence to a protein through the action of a ribosome molecule.

A genome is fundamentally a long chain, or polymer, made up of pairs of
nucleotide monomers, of which there are four (adenine, cytosine, guanine, and
thymine). It is productive, however, to consider groupings of nucleotides
according to their function. If the entire genome is a sequence of sentences,
and the amino acids are letters, then by analogy, a \emph{biological part} is a
word. Continuing the analogy, just as words come in several classes, or parts of
speech, so do biological parts: promoters, coding sequences, binding sites,
terminators. Furthermore just as a sentence may be well formed or ungrammatical
according to what kinds of words come in what order, the same is also true of a
genome~\citep{pedersen_towards_2009} --- particularly the kind of synthetic
genome that one might wish to construct.

\begin{figure}[h]
  \includegraphics[width=\textwidth]{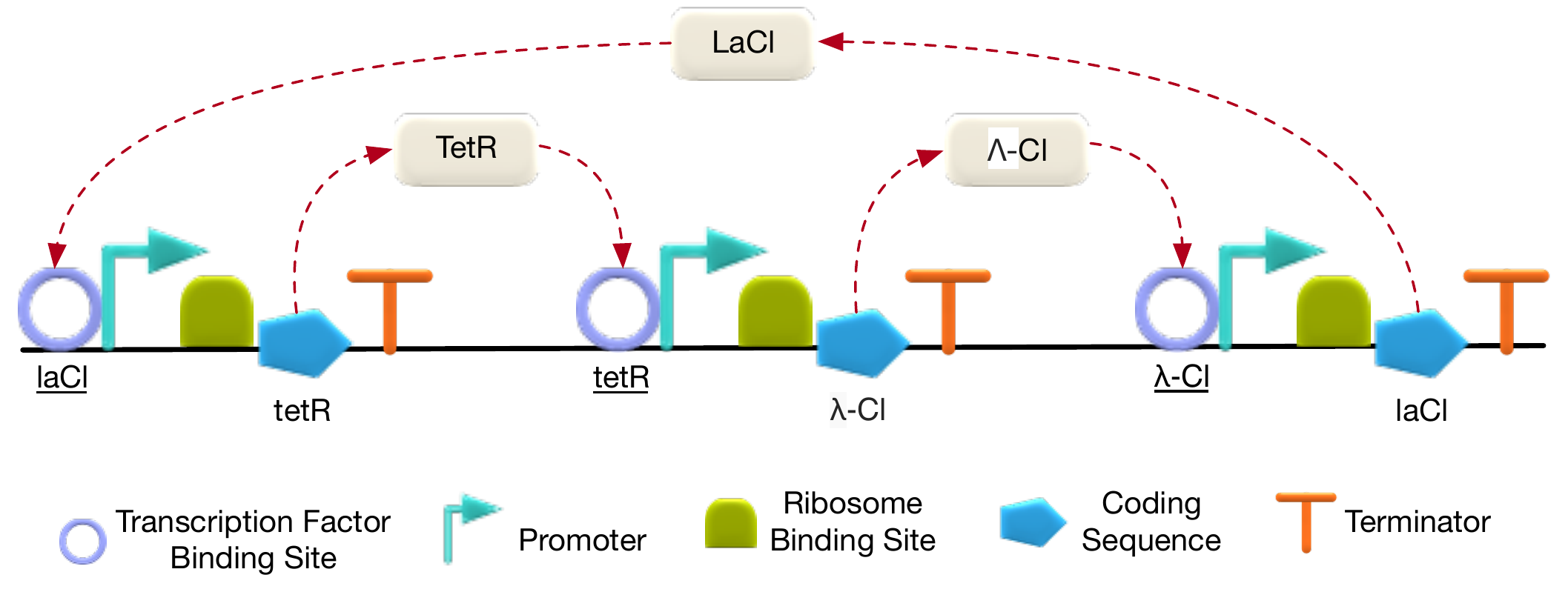}
  \caption{An example genetic circuit: the Elowitz repressilator. It is a
    negative feedback oscillator. The circuit itself is arranged linearly and
    protein-operator interactions are represented by arrows and dotted lines.}
  \label{fig:repressilator}
\end{figure}
To understand the basic features of this grammar, and the functioning of
a genetic circuit, consider an example both simple and famous, the Elowitz
repressilator~\citep{elowitz_synthetic_2000}, shown in
Figure~\ref{fig:repressilator}. It consists of three sentences. A sentence
begins with a binding site for a transcription factor and a promoter ---
together, an operator. If the
conditions of the binding site are satisfied, by being bound or not by some
specific protein, then the promoter is activated. If the promoter is active, an
RNA polymerase may bind it, from which it will slide across
and transcribe a binding site for a ribosome and coding sequence that will
result eventually in the production of a particular protein. Finally a
terminator ends the transcription process and the ribosome falls off.

In the case of this circuit, each operator is active by default. Each sentence
in turn will produce a protein that will bind to the next operator, deactivating
it. As the proteins eventually degrade (not shown), the operators reactivate.
This interaction sets up an oscillation that can be observed by measuring the
concentrations of the three proteins. Also not shown here is the intermediary
stage --- the output of transcription is RNA, and it is the translation of RNA
that actually produces the proteins.

\subsection{Rule-based Modelling of Genetic Processes}
\label{sec:kappa-illustration}

It might seem that the processes of transcription and translation, and the
interactions of proteins with operators could be represented by reactions. After
all, these are chemicals, and certain input chemicals (nucleic acids, enzymes
like RNA polymerase and ribosomes, and proteins) interact and produce other, new
chemicals (more proteins). It turns out that this approach does not scale well.

It is easy to see why. Suppose the input to the process is a large DNA module,
the entire genome, together with a starting population of RNA polymerase and
ribosomes. The first interaction might be an RNA polymerase binding somewhere on
the DNA. In the example above, there are three different places that this can
happen which result in six possible different output molecules of the DNA
carrying a bound RNA polymerase. That is only the first step, now the RNA
polymerase must move to the right and begin transcription. This next step now
requires six different reactions to describe it, one for each flavour of
RNAp-bound DNA.

To solve this problem of needing combinatorially many reactions to describe
essentially the same process, a generalisation of reactions called \emph{rules}
are used~\citep{hlavacek2003complexity,danos2004formal,danos2008rule}. In this
representation, \emph{agents} correspond to reagents and they can have slots or
\emph{sites} that can be bound, or not. They can also have internal state.
Unlike reactions which have no preconditions apart from the presence of the
reagents, with rules, a configuration of the sites --- bound in a particular
way, bound in some way, unbound, or unspecified --- is a precondition for the
application of the rule. A rule may re-arrange the bonds, creating or destroying
them, without the need to invent new agents in order to represent different
configurations of a given set of molecules.

The reader should note that the word \emph{rule} is used in two distinct senses
in this article. The first is as we have just described. The second is in the
sense of \emph{inference rule} as used in logic and in particular the way in
which we deduce executable rule-based models from their declarative
representations in RDF. 

\subsection{The $\kappa$ Language}
To briefly illustrate the essentials of rule-based modelling we will use the
language of the Kappa simulation software, KaSim~\citep{krivine_kasim_2017}. An
agent declaration and rule expressing the formation of a polymer might look
like,
\begin{lstkappa}

'binding' A(u), A(d) -> A(u!1), A(d!1) @k
\end{lstkappa}
We can gloss this as an agent with two sites, \lstinline{u} and \lstinline{d}
for upstream and downstream, and a rule. The rule concerns two agent patterns
one of which has an unbound upstream site, and the other an unbound downstream
site, and the action of the rule is to bind them, the notation \lstinline{!1}
denoting the bond. This process happens at some rate, with the rule applied
\lstinline{k} times per unit time on average.

The state of the other site of each agent is left unspecified, so implicit
in this rule is the possibility that either or both the agents may already be
bound to others and so part of arbitrarily long chains.
In other words this expression covers not only two monomers joining together but
an $n$-mer and an $m$-mer for arbitrary $n$ and $m$.
This is the essence of the expressive advantage that rule-based modelling
provides. 
To express a similar concept using a reaction network would in fact require
infinitely many reagents for every possible $n$ (and $m$) and infinitely many
reactions for every possible combination.

\subsection{The $\kappa$ BioBrick Framework}
In order to apply the general rule-based modelling approach to the specific
question of modelling genetic circuits, we adopt the methodology of
the~\ac{KBBF}~\citep{marchisio_kappa_2015}.
This framework provides a set of rules that describe the transcription and
translation of DNA parts and a modeller provides rules for the interactions of
gene products.

\ac{KBBF} focused on the composition of DNA-based parts without enforcing rules
about the interactions involving gene products such as proteins. Composition of
the parts into complex circuits was only manually demonstrated; the decision to
choose and group rule sets to model individual parts was left entirely to the
modeller.

\subsection{Biological Parts and Annotation}
For efficiency, and economy of representation, a computational model should
include minimum information necessary 
for simulation. However, in order to use
these models in an automated design process, additional metadata, or
annotations, about the meaning of different modelling entities is
needed~\citep{neal2014reappraisal}. 

One might wish to draw specific parts from a database such as the Virtual Parts
Repository~\citep{misirli_composable_2014}.
These models are annotated with machine-readable metadata to facilitate their
combination into larger models and thereby rendering large parts of the
design space for biological systems amenable to automation.
Furthermore, Myers and his colleagues have used annotations to derive simulatable models from
descriptions of genetic circuits~\citep{roehner2015} and vice versa~\citep{nguyen2016}.
These previous works use reaction-based models represented in the Systems
Biology Markup Language (SBML)~\citep{hucka2003} and so inherit the poor
scaling properties of that method.

Annotation in this setting means machine-readable descriptions of entities of
biological interest. This is done with statements, triples of the form
(subject, predicate, object) according Semantic Web
standards~\citep{cyganiak_rdf_2014,van_harmelen_owl_2004}. Entities are
identified with \acp{URI}~\citep{masinter_rfc3896_2005}. This provides the dual
benefit of globally unique identifiers for entities and a built-in mechanism for
retrieving more information about them providing that some care is taken to
publish data according to best
practises~\citep{hyland_best_2014,sauermann_cool_2011}. Large bodies of such
information about biologically relevant information are published on the
Web~\citep{ashburner_gene_2000,consortium_universal_2008} and the use of Semantic
Web standards for annotating our models allows us to refer, to say that an
entity in a model description corresponds to a real world protein, or gene
sequence or the like.

The Semantic Web also affords us a technical advantage: inference rules.
These can be either explicit as in
Notation3~\citep{berners-lee_n3logic:_2008,berners-lee_notation3_2011} or
implicit as in OWL Description
Logics~\citep{horrocks_owl:_2005,brickley_rdf_2014}. In either case
this facility makes it possible, given a set of statements, to derive new
statements according to rules. We use this to improve the ergonomics of
our high-level language: while the compiler itself will make use, internally, of
a large amount of information, we do not expect the user to supply it all in
painstaking detail. Rather we allow the user to specify the minimum possible and
provide rules to derive the necessary detail. This gives both economy of
representation for the high-level model description and flexibility for
the different implementations.

\section{A Language for Synthetic Gene Circuits}
\label{sec:gcdl}

To facilitate the \textit{in silico} evaluation of potential synthetic gene
circuits, a library of descriptions of genetic parts, together with their
modular models is suggested
in~\citep{cooling_standard_2010,misirli_composable_2014}.
These parts are
intended to be large enough to have a particular meaning or function (i.e.
larger than individual base pairs) but not so large that they lack the
flexibility to be easily recombined (i.e. entire genes).
Thus we are concerned with coding sequences for particular proteins, promoters
that, when activated, start the transcription process, operators that activate
or suppress promoters according to whether they are bound or not by a given
protein, and a small number of other objects.
A sequence of these objects is a genetic circuit, and our goal is to have a good
language for describing such sequences.

\subsection{Desired Language Features}
Our desiderata for a high-level representation of a genetic circuit are as follows,
\begin{enumerate}
\item sufficiency, there should be enough information to derive
  executable code for the circuit,
\item identifiability, it should be possible to determine to which biological
  entities (DNA sequences, proteins) the representation refers,
\item extensibility, it should be straightforward to add information or constructs that
  are not presently foreseen,
\item generality, there should be no requirement that information about
  biological parts comes from any particular set or source, and
\item concision, there should be a minimum of extraneous detail or syntax.
\end{enumerate}
The third and fourth are readily accomplished by using \acs{RDF} as the
underlying data model.
The \emph{open world} presumption~\citep{drummond_open_2006} means that adding
information as necessary is straightforward.
The use of \acs{URI}s~\citep{masinter_rfc3896_2005} which can be dereferenced
to obtain the required information means that information from different
web-accessible databases can be obtained, mixed and matched as desired.
The second desideratum is assisted by the use of \acs{URI}s, albeit
with some well-known caveats~\citep{halpin_when_2010}.

The first and last of the desired features are, therefore, primarily what
concerns us in this paper. To begin with, we suggest (but do not require) the
use of Turtle~\citep{prudhommeaux_rdf_2014} or indeed
Notation3~\citep{berners-lee_rdf_2005} as the concrete surface syntax for
writing models. This goes some way towards a representation that is
intelligible by humans. Even then, we aim to minimise what needs to be
written and we do this using inference rules --- if a needed fact can be derived from
the model under the provided rule-set, it is unnecessary to write it explicitly
in the model. Indeed it may even be undesirable to do so since it is a possible
source of errors such as being correct in the context of some output types and
incorrect in others. So, we aim for a minimal, yet complete under the inference
rules, description of the model.

\subsection{Model Description}

To illustrate the syntax of the high-level language, we return to the 
Elowitz repressilator of Figure~\ref{fig:repressilator}.
Figure~\ref{fig:repressilator-model} shows a description of this circuit
in the \ac{GCDL}.
It is written in Turtle and identified as a model using terms from the
\ac{RBMO} that we previously defined~\citep{misirli_annotation_2015}.
\begin{figure}
  \begin{lstturtle}
  ## Model declaration
  :m a rbmo:Model;
    ## bibliographic metadata
    dct:title "The Elowitz repressilator constructed from BioBrick parts";
    dct:description "Representation of the Elowitz repressilator given in the Kappa BioBricks Framework book chapter";
    rdfs:seeAlso <http://link.springer.com/protocol/10.1007/978-1-4939-1878-2_6>;
    gcc:prefix <http://id.inf.ed.ac.uk/rbm/examples/repressilator#>;
    ## include the host environment
    gcc:include <.../host.ka>;
    ## initialisations
    gcc:init
      [ rbmo:agent :RNAp; gcc:value 700 ],
      [ rbmo:agent :Ribosome; gcc:value 1000 ];
    ## The circuit itself, a list of parts
    gcc:linear (
      :R0040o :R0040p :B0034a :C0051 :B0011a
      :R0051o :R0051p :B0034b :C0012 :B0011b
      :R0010o :R0010p :B0034c :C0040 :B0011c
    ).
  \end{lstturtle}
  \caption{Example model for a synthetic gene circuit, Elowitz' repressilator.}
  \label{fig:repressilator-model}
\end{figure}
Some bibliographic metadata is included, using the standard Dublin
Core~\citep{kunze_rfc5013_2007} vocabulary, as well as a generic pointer
(\term{rdfs:seeAlso}) to a publication about this model. 

New terms introduced in this paper have the prefix \term{gcc} which can be read
as the ``Genetic Circuit Compiler'' vocabulary. The term \term{gcc:prefix} is
necessary in every model, it instructs the compiler that any entities that it
creates should be created under the given prefix. Ultimately annotated rules
will be generated for the low-level representation and the annotated entities
require names. To give them names, a namespace is required and this is how it is
provided.

Next there is a \term{gcc:include} statement. This is a facility for including
extra information in the low-level language. Extra information typically means
rules for protein-protein interactions which are beyond the scope of the current
work and as such it is simply supplied as a program fragment in the output
language. This corresponds roughly to dropping to assembly or machine language
to perform a specialised task when programming a computer in a high-level
language like C.

There follows initialisation for specific variables. In this case these are the
copy numbers for RNA polymerase molecules and ribosomes. These are denoted using
\term{rbmo:agent} because of our choice to support rule-based modelling for
greater generality than reaction-based methods. 

Finally, the circuit itself is specified. The argument, or object is an
\term{rdf:List} which simply contains identifiers for the parts, in order.

The circuit itself is now defined. However at this juncture, we simply have a
list of parts without having specified what they are. To obtain a working model,
we need more.

\subsection{A Part Description}

A simple example of a part description is shown in
Figure~\ref{fig:coding-sequence-part}. This is a coding sequence, as is clear
from the type annotation on the part. It codes for a particular protein,
specified with \term{gcc:protein}. This term is specific to proteins because
under normal circumstances other kinds of part do not code for proteins.
\begin{figure}[t]
  \begin{lstturtle}
  :P0010 a gcc:Protein;
      bqbiol:is uniprot:P03023;
      skos:prefLabel "P0010";
      rdfs:label "LacI".

  :C0012 a gcc:CodingSequence;
      gcc:label "Coding sequence for LacI";
      gcc:part "C0012";
      gcc:protein :P0010.
  \end{lstturtle}
  \caption{A coding sequence part description from the repressilator model.
    Notice how the coding sequence is linked to the protein that it codes for.}
  \label{fig:coding-sequence-part}
\end{figure}
The protein itself is also described, mainly so that it can be given a label,
which is done using the term \term{skos:prefLabel} from the
\ac{SKOS}~\citep{miles_skos_2005} vocabulary.
This detail is important. Because the output language will not typically permit
the use of \acs{URI}s as identifiers, the preferred label is used as the
identifier for the protein in the output representation.
This same is true of the value given for \term{gcc:part}. That term implies the
corresponding \term{skos:prefLabel} by means of 
inference rules supplied with the \term{gcc} vocabulary.

Importantly, and following the practice in our previous paper on rule
annotation~\citep{misirli_annotation_2015}, a weak identity assertion is made
with identifiers in external databases for the parts.
This uses \term{bqbiol:is} instead of \term{owl:sameAs}
because the strong Leibniz identity semantics of the latter can yield unwanted
inferences when terms are not used perfectly rigorously~\citep{halpin_when_2010}.
This weaker identify assertion permits the identification of the \term{:P0010} in
the example with the identifier for the protein in the well-known
UniProt~\citep{consortium_universal_2008} database.

A part will typically have associated rate data. These rates characterise the
interaction of the part with RNA polymerase  and the production of the
corresponding RNA and the RNA's interaction with a ribosome to produce
proteins. This information comes primarily from experiment and is the main
reason why it is important to have accessible databases or repositories of part
specifications.  
In this case, no rates are explicitly specified here, so they will take on
default values.

\subsection{A More Complex Part Description}
A more involved example demonstrating how an operator-promoter combination is
encoded is shown in Figure~\ref{fig:repressible-promoter-parts}. Here we have an
operator with the rates for binding and unbinding of the transcription factor
specified explicitly. If the operator is bound by the transcription factor, the
neighbouring promoter is repressed --- an RNA polymerase will not be able to bind.
By contrast if the operator is unbound, the promoter will
accept binding of RNA polymerase easily and frequently.

The transcription factor itself is specified by using
\term{gcc:transcriptionFactor} to refer to the protein that will turn the
operator on or off. Like \term{gcc:protein} for coding sequences,
the term is unique to operators.
\begin{figure}
  \begin{lstturtle}
  :R0040o a gcc:Operator;
    rdfs:label "TetR activated operator";
    gcc:part "R0040o";
    gcc:transcriptionFactor :P0040;
    gcc:transcriptionFactorBindingRate 0.01;
    gcc:transcriptionFactorUnbindingRate 0.01.

  :R0040p a gcc:Promoter;
    rdfs:label "TetR repressible promoter";
    gcc:part "R0040p";
    gcc:rnapBindingRate
      [
        gcc:upstream ([ a rbmo:BoundState;
                         rbmo:stateOf :R0040o ]);
        gcc:value 7e-7
      ], [
        gcc:upstream ([ a rbmo:UnboundState;
                         rbmo:stateOf :R0040o ]);
        gcc:value 0.0007
      ].
  \end{lstturtle}
  \caption{An operator and promoter from the repressilator model. The binding
    rates for the promoter depend on the state of the adjacent operator.}
  \label{fig:repressible-promoter-parts}
\end{figure}

The promoter itself comes next and it is the most complex part to
specify.
Because the rate for binding of RNA polymerase depends on the state of
the operator, two rates must be specified.
States of the nearby parts are specified using the \term{rbmo} vocabulary which
makes available the full range of expressiveness for rule-based output
languages. 
For generality, a list of parts, upstream or downstream on the DNA strand may be
specified along with their states.
This enables a promoter to be controlled by two or more operators.
The rate itself in this
case is given with \term{gcc:value} for each case.

\subsection{Host and Protein-Protein Interactions}
\begin{figure}[t]
  \begin{subfigure}{\textwidth}
    \includegraphics[width=\textwidth]{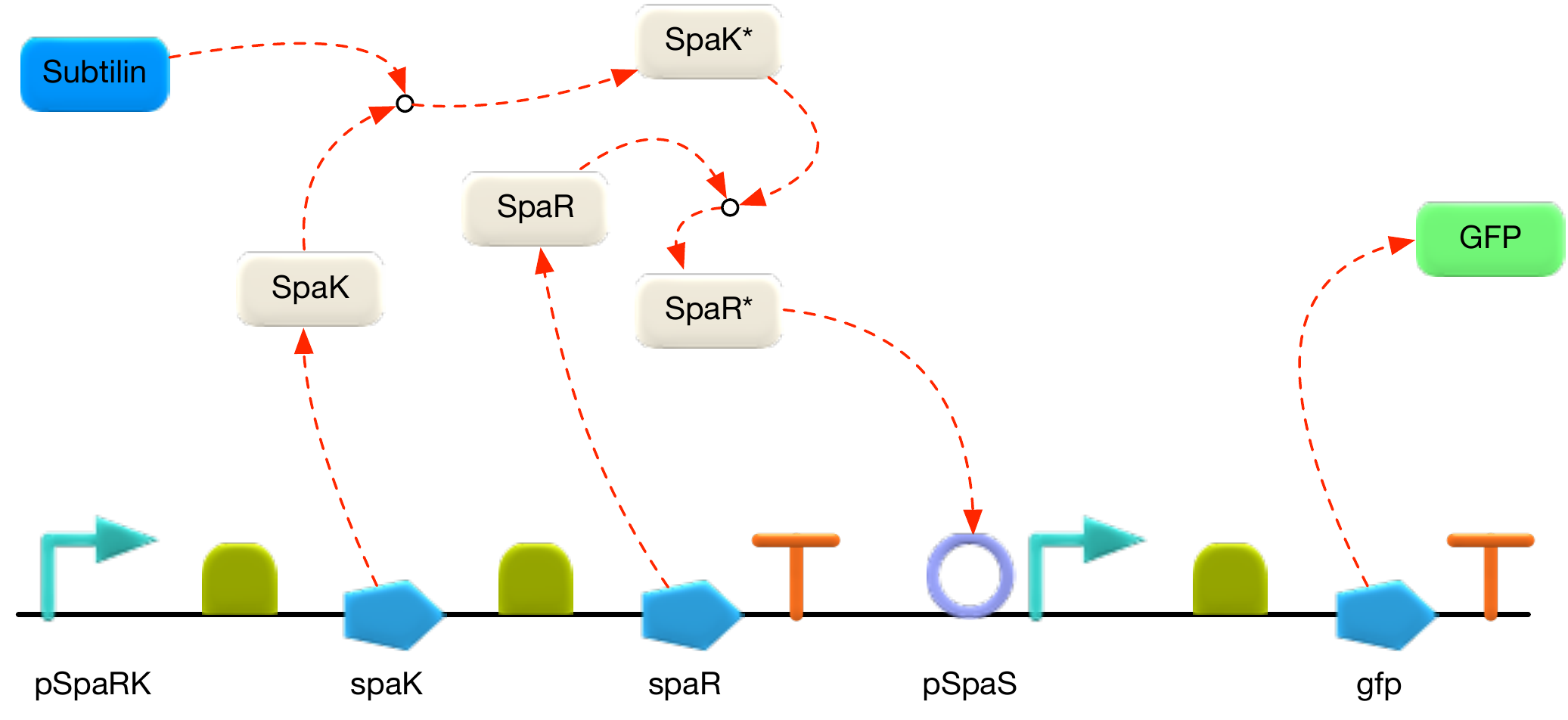}
    \caption{Diagram of the genetic circuit.}
    \label{fig:subtilin-receiver}
  \end{subfigure}
  \begin{subfigure}{\textwidth}
    \begin{lstturtle}
  :m a rbmo:Model;
      dct:title "Subtilin Receiver Two-Component System";
      gcc:include <.../subtilin-host.ka>;
      gcc:linear (
          :pSpaRK :RBSa :spaK :RBSb :spaR :Ta
          :pSpaS  :RBSc :gfp :Tb
      ).
    \end{lstturtle}
    \caption{Corresponding semantic model.}
    \label{fig:subtilin-receiver-model}
  \end{subfigure}
  \caption{Representations of the Subtilin Receiver model.}
\end{figure}

The language can also support protein--protein interactions.
To see why these are useful, consider an example from the engineering of a
bacterial communication system where the subtilin molecule is used to control
population level dynamics.
Cells has the receiver device
~\citep{bongers_development_2005,misirli_composable_2014} to sense the
existence of subtilin, and the reporter device to initiate downstream cellular
processes (Figures~\ref{fig:subtilin-receiver}
and~\ref{fig:subtilin-receiver-model}).
In the subtilin receiver, the interactions among the proteins produced by
translation and the operator-promoters are mediated by a cascade reaction
initiated by the subtilin molecule.
Subtilin combines to phosphorylate the \emph{SpaK}
protein, which in turn phosphorylates the \emph{SpaR} protein that finally binds
to the promoter that controls the emission of a fluorescent green protein.

While the genetic circuit itself can be described in a similar manner to the
previous repressilator example, the protein--protein interactions cannot.
We do not attempt here to model these interactions in the \ac{GCDL} though a
future extension could do so.
Instead we simply allow for inclusion of the relevant program, as a file in the
output language (in this case $\kappa$-language).
It is possible to supply arbitrary code in the
low-level language using the \term{gcc:include} term.
This facility makes it feasible to represent such genetic circuits which depend
strongly on the host environment in order to operate.

\subsection{Protein Fusion}
It is also worth noting that this example illustrates that in the high-level
language it is immediately possible to represent devices that produce chains of
proteins.
This is known as protein fusion and is interesting for some
applications~\citep{yu2015synthetic}. 
A chain of proteins is produced by adding adjacent (and appropriate) coding
sequences.
It is enough to simply list the coding sequences in the circuit; 
nothing else need be done.

\subsection{Other Parts}
The descriptions for the other kinds of biological parts, terminators, coding
sequences, follow a similar pattern.
There are terms for specifying the rates for the rules in which they participate, and
a few specialised terms according to the function of the specific part.
It is possible to find the available terms out by inspecting the \term{gcc}
vocabulary included in Appendix~\ref{sec:gcc-vocabulary}.

\section{Output Representations}
\label{sec:ruleoutput}

We now consider the form of the output representation. By using different
templates, the compiler can produce output in different languages. We focus
on rule-based representations here and use the language of the KaSim
simulator~\citep{krivine_kasim_2017} for concrete illustration as it is widely
adopted for stochastic simulation of rule-based
models~\citep{marchisio_kappa_2015}.
The rule-based modelling approach is merely outlined here and follows that
used in \ac{KBBF}~\citep{marchisio_kappa_2015} closely. We stress that the
though output as  executable program in the KaSim language is
demonstrated here, alternative rule-based representations like 
BioNetGen are equally possible as are
descriptions in a language like \ac{SBOL} as input to an
experimental process in the laboratory.

\subsection{Generic Agents}

The behaviour of each kind of genetic part can be specified
with rules, examples of which are given below.
Fundamentally these rules operate on representations of DNA, RNA and proteins.
Since each part can be linearly adjacent to others,
there must be sites to stand for this linkage.
These will be called \lstinline{us} and \lstinline{ds} for ``upstream'' and
``downstream'' respectively.
There is also a need for a site to stand for the binding of
protein or RNA polymerase to DNA, or the ribosome to RNA.
This will be called \lstinline{bs} for ``binding site''.

We immediately arrive at a modelling choice: the specific part, for example an
operator to which the Lac repressor binds, could be represented as distinct kind
of agent with DNA, RNA and protein variants (Figure~\ref{fig:partdualdistinct}) or it could be represented as a
label or tag on a generic DNA, RNA and protein agents (Figure
\ref{fig:partdualgeneric}).
\begin{figure}
  \begin{subfigure}{\linewidth}
    \begin{lstkappa}
    \end{lstkappa}
    \caption{Distinct agents for each variant of part.}
    \label{fig:partdualdistinct}
  \end{subfigure}
  \begin{subfigure}{\linewidth}
    \begin{lstkappa}
    \end{lstkappa}
    \caption{Generic agents for each variant with part indicated by the
      \lstinline{type} site.}
    \label{fig:partdualgeneric}
  \end{subfigure}
  \caption{Dual representations of parts as agents.}
  \label{fig:partdual}
\end{figure}
We choose the latter because not only does it remove the need for having a large
number of agents and inventing names for each DNA and RNA variant, but it
greatly simplifies the rules. As we shall see the generic representation means
that rules can easily be written where it only matters that a part is adjacent
to \emph{some} other part without specifying which one in particular. This is
simply done by not specifying the \lstinline{type}
site. This is not possible
with distinct agents because the Kappa language does not allow for unspecified or
wild card agents.

\begin{figure}
  \begin{lstkappa}
  \end{lstkappa}
  \caption{RNA polymerase and ribosome agents.}
  \label{fig:straddle}
\end{figure}
These constructs, with their upstream and downstream linkages are enough to form
the ``rails'' along which transcription and translation happen but we still
require agents to join these together, namely RNA polymerase and the ribosome.
These agents have two sites, one for each rail that they straddle
(Figure~\ref{fig:straddle}).

\subsection{Unbinding Rules}

\begin{figure}
  \begin{subfigure}[t]{0.5\linewidth}
    \begin{centering}
      \includegraphics[width=0.6\linewidth]{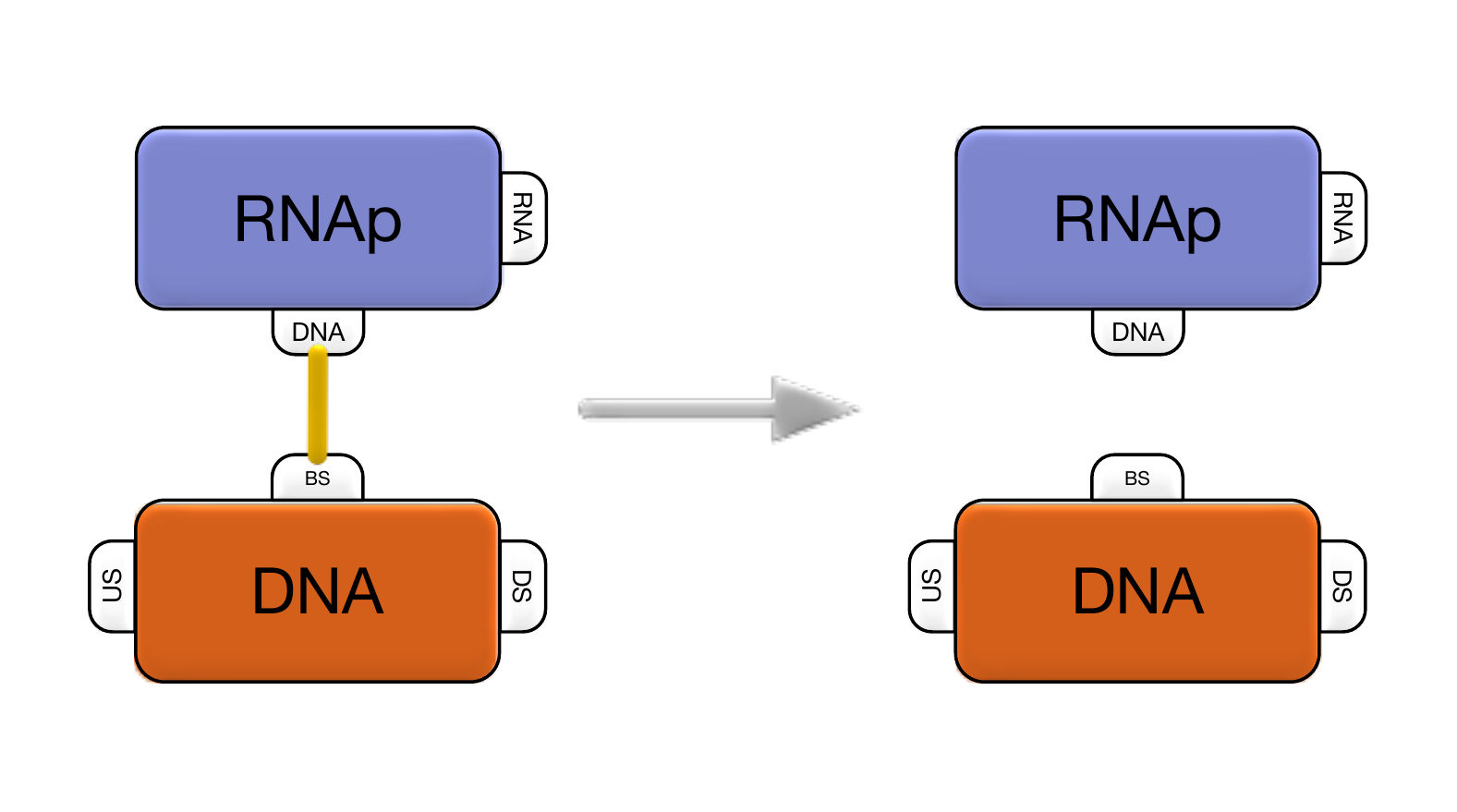}\\
      \includegraphics[width=\linewidth]{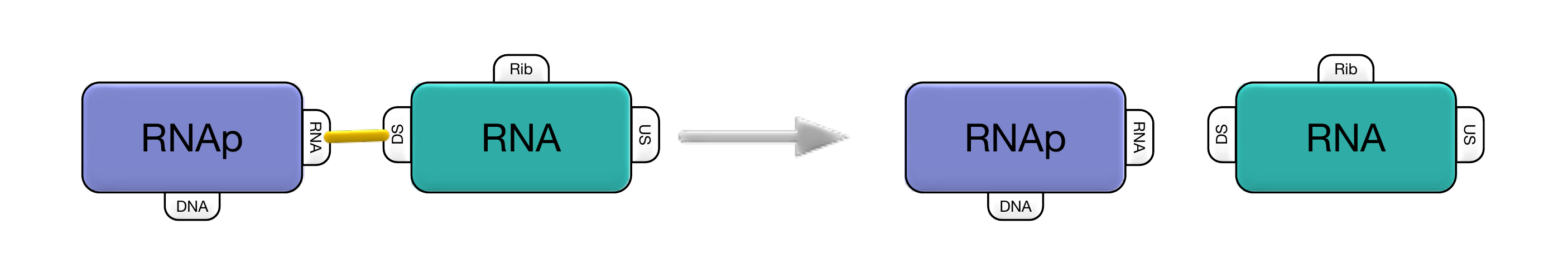}
    \end{centering}
  \end{subfigure}
  \begin{subfigure}{0.5\linewidth}
    \begin{lstkappa}
'transcription termination' \
  DNA(bs!1), RNAp(dna!1) -> DNA(bs), RNAp(dna) @k

'translation termination' \
  RNA(ds!1), RNAp(rna!1) -> RNA(ds), RNAp(rna) @k
\end{lstkappa}
\end{subfigure}
  \caption{Termination rules: transcription and translation}
  \label{fig:transcription-termination}
\end{figure}

To understand how this works in practice, consider the simplest kind of rule,
the unbinding rule. Those for transcription and translation are shown in
Figure~\ref{fig:transcription-termination}.
This does not yet use any of the features that motivated our choice of agent
representation, but does already show the ``don't care, don't write'' way of the
KaSim dialect of Kappa: those sites that are not necessary for the operation of
the rule do not appear. This brevity is a great boon.

An unbinding rule of the same form exists for each DNA part. Particularly
significant among these is the unbinding of a protein from an operator. 


\subsubsection{Binding Rules with Context}

\begin{figure}
  \begin{subfigure}[b]{0.5\textwidth}
    \begin{center}
      \includegraphics[width=\linewidth]{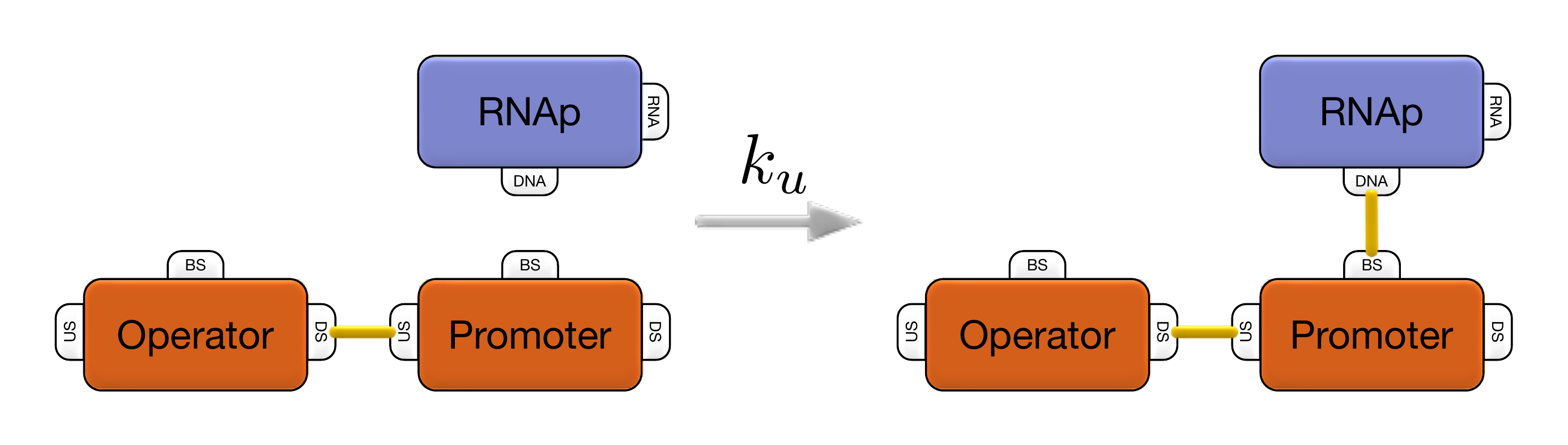}
      \par\vspace{\baselineskip}
      \includegraphics[width=\linewidth]{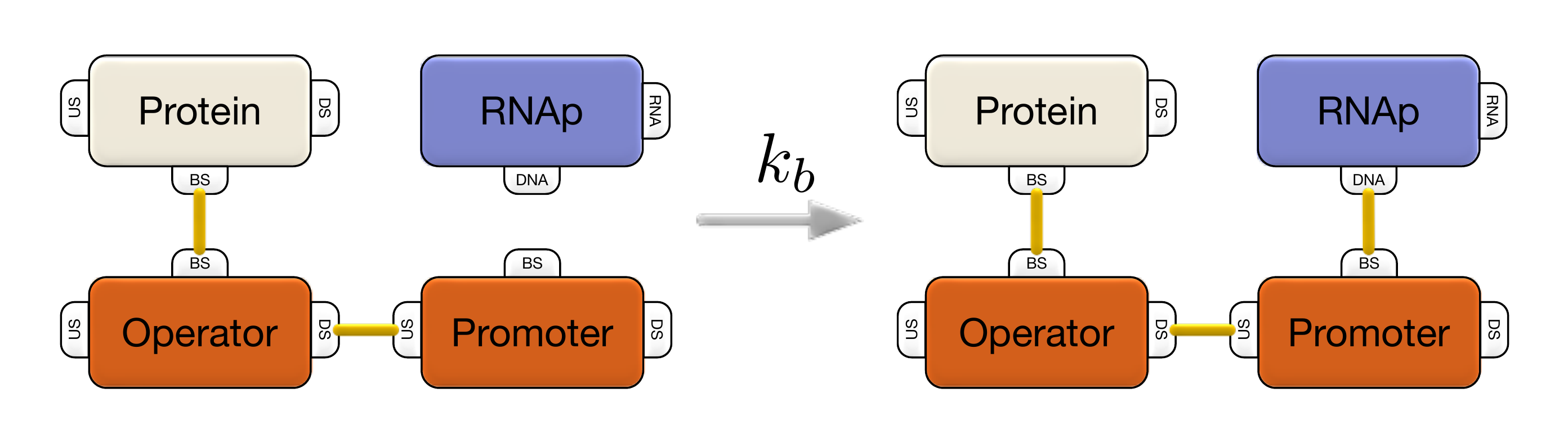}
    \end{center}
  \end{subfigure}
  \begin{subfigure}[b]{0.5\textwidth}
  \begin{lstkappa}
  'RNAp-binding-unbound' \
      DNA(type~operator, ds!1, bs), \
      DNA(type~promoter, us!1, bs), \
      RNAp(dna) \
  -> \
      DNA(type~operator, ds!1, bs), \
      DNA(type~promoter, us!1, bs!2), \
      RNAp(dna!2) \
  @k_u

  'RNAp-binding-bound' \
      DNA(type~operator, ds!1, bs!_), \
      DNA(type~promoter, us!1, bs), \
      RNAp(dna) \
  -> \
      DNA(type~operator, ds!1, bs!_), \
      DNA(type~promoter, us!1, bs!2), \
      RNAp(dna!2) \
  @k_b
\end{lstkappa}
\end{subfigure}
  \caption{Binding of RNA polymerase to a promoter with different rates, $k_u$
    and $k_b$ according to context given by operator state.}
  \label{fig:promoter-activation}
\end{figure}

The simplest kind of binding rule is just the same as unbinding with the
direction of the arrow reversed. Such rules appear for the initiation of
translation --- the binding of a ribosome onto a ribosome binding site --- as
well as for the activation of an operator. These are not reproduced here.
Instead, we consider binding rules with context, as in
Figure~\ref{fig:promoter-activation}.

The explicit context, with the operator adjacent to the promoter being bound to
a protein, or not, allows for the modelling of inducible or repressible promoter
architectures. The transcription process begins with the binding of RNA
polymerase and the rate at which this happens depends on the state of the
operators as illustrated in Figure~\ref{fig:promoter-activation}. This is the
simple case with only one operator but there is no restriction on the number of
operators; we allow for upstream and downstream context of arbitrary size.

This example is illustrative in that rules are posed in terms of a ``main'' part
that becomes bound or unbound and in principle it is possible to provide
arbitrary amounts of context for \emph{any} rule. This is supported by the
low-level language here, but however it is only implemented in the compiler for
the particular family of rules depicted in Figure~\ref{fig:promoter-activation},
the activation of promoters through the binding of RNA polymerase. This is
sufficient for models involving complex promoter architectures, but an extension
allowing for context everywhere is not difficult.

\subsubsection{Sliding Rules}

\begin{figure}
  \begin{subfigure}[b]{0.5\textwidth}
    \centering
    \includegraphics[width=\linewidth]{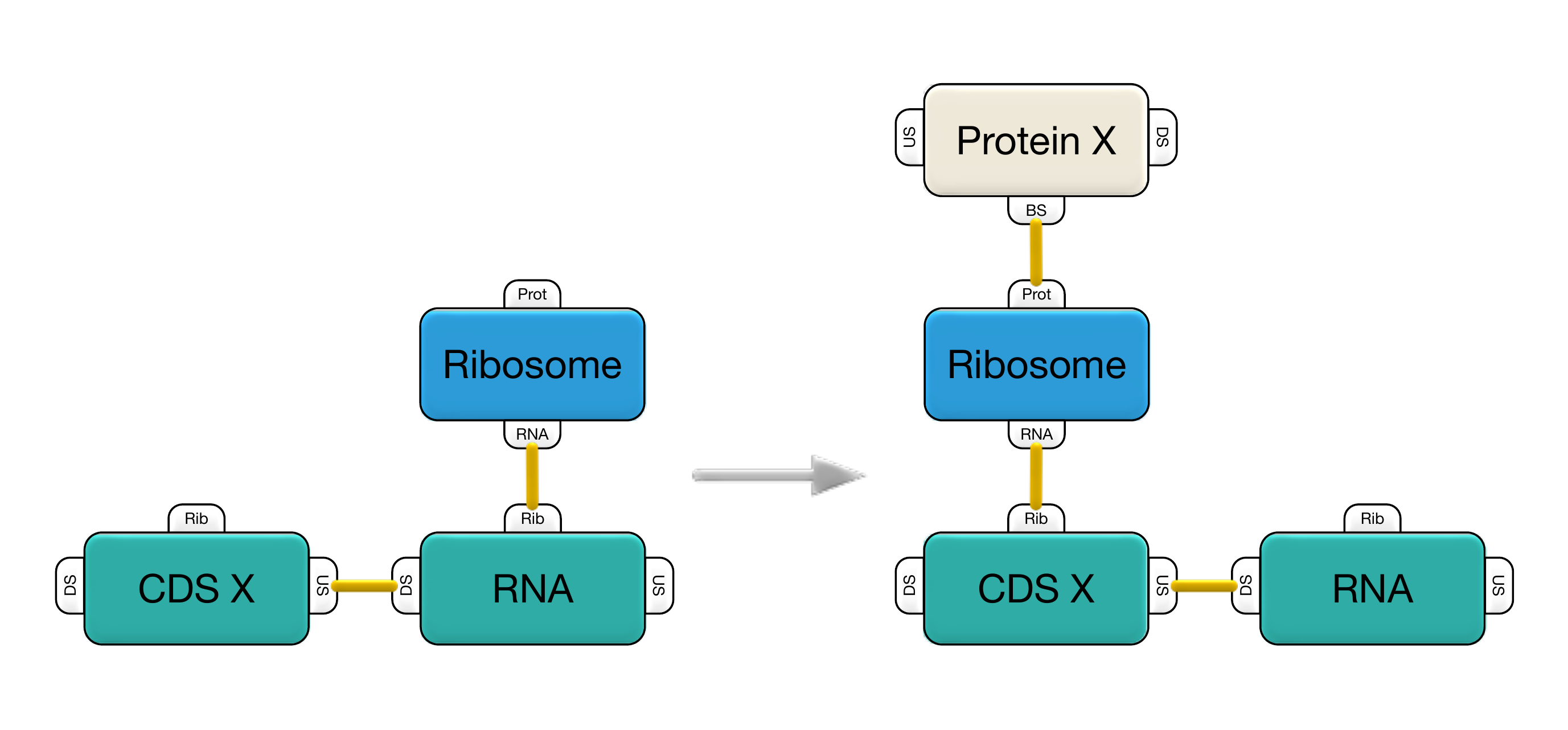}
  \end{subfigure}
  \begin{subfigure}[b]{0.5\textwidth}
  \begin{lstkappa}
  'coding-sequence-translation' \
      RNA(type~X, us!2, bs), \
      RNA(ds!2, bs!1), \
      Ribosome(rna!1, protein) \
  -> \
      RNA(type~X, us!2, bs!1), \
      RNA(ds!2, bs), \
      Ribosome(rna!1, protein!3), \
      P(type~X, bs!3) \
  @k
\end{lstkappa}
\end{subfigure}
  \caption{Translation of a coding sequence to produce a protein.}
  \label{fig:coding-sequence-translation}
\end{figure}

In some sense the real work of modelling the transcription and translation
machinery is done with sliding rules.
Figure~\ref{fig:coding-sequence-translation} shows how this works for the
creation of a protein from a coding sequence. This is our
first example of a rule where though the adjacent part figures explicitly in the
rule, its \emph{type} does not. It is sufficient to know that it is a piece of
RNA.
In this case, two pieces of RNA are involved, the part that is central to this
rule corresponds to the coding sequence for $X$. It is adjacent to another piece
of RNA, and the ribosome slides from one to the other (to the left, where
sliding on DNA happens, as we will see next, to the right) and in the process,
emits a protein of type $X$.

\begin{figure}
  \begin{subfigure}{0.65\textwidth}
    \centering
    \includegraphics[width=\linewidth]{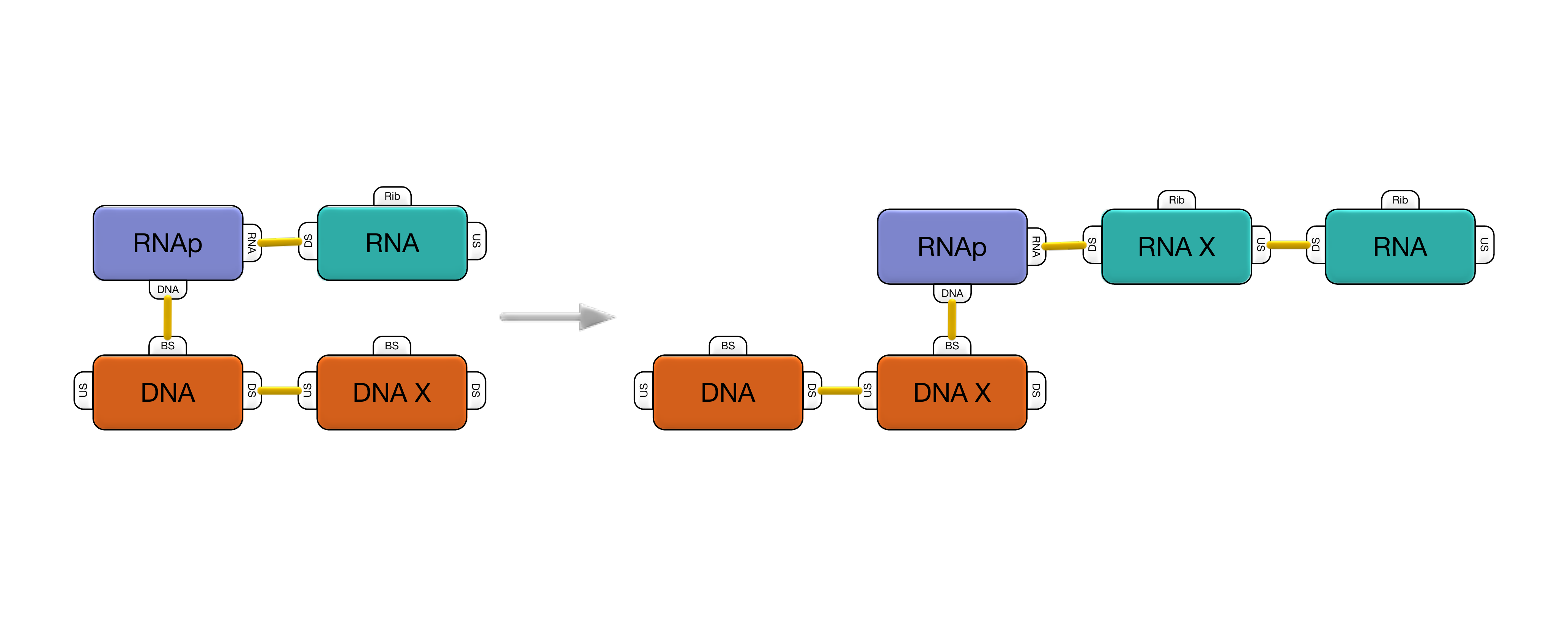}
  \end{subfigure}
  \begin{subfigure}{0.35\textwidth}
  \begin{lstkappa}
  'transcription-elongation' \
      DNA(ds!2, bs!1), \
      DNA(type~X, us!2, bs), \
      RNAp(dna!1, rna!3), \
      RNA(ds!3) \
  -> \
      DNA(ds!2, bs), \
      DNA(type~X, us!2, bs!1), \
      RNAp(dna!1, rna!3), \
      RNA(ds!4), \
      RNA(type~X, us!4, ds!3, bs) \
  @k
\end{lstkappa}
\end{subfigure}
  \caption{Transcription, production of an RNA sequence from DNA}
  \label{fig:transcription-elongation}
\end{figure}

A somewhat more complicated sliding rule is used to implement transcription, as
shown in Figure~\ref{fig:transcription-elongation}. This shares the feature of
the translation rule above where there is a part that is central to this rule,
part $X$, and there is an adjacent part whose type does not matter.
Here, the RNA polymerase starts off bound to the adjacent DNA part, whose type
does not matter and so is not specified, and slides onto the central part of
type $X$. In the process, an RNA part of type $X$ is inserted into the growing
chain.

Other rules are necessary, of course. The rule in
Figure~\ref{fig:transcription-elongation}, for example, cannot operate without
a piece RNA bound to the polymerase. Chains of RNA cannot be produced before the
first link has been added. The rule that does that is exactly
analogous to that of Figure~\ref{fig:coding-sequence-translation}. And similarly
in the other direction, there is a rule to produce protein chains where a
protein already exists and a coding sequence is slid across. This is almost
identical to making an RNA chain. All of the other core rules are simply
variations on those given above.

\section{Genetic Circuit Compiler}
\label{sec:gcc}

Having described the source (high-level) and target (low-level) languages in
some detail, we now briefly sketch our implementation of the compiler that
translates between them. Many compiler implementations are possible; ours
innovatively combines the logical inference that is native to the semantic web
with the use of templates to generate the target program. The overall
information flow through the compiler is illustrated in
Figure~\ref{fig:compiler-flow}.

Our strategy is to first gather all the input statements and background facts
that are asserted by the various vocabularies in use. In the first inference
step, standard RDF rules are then used to make available consequent facts that
will be needed to produce the ultimate result. The result is a program in a
language such as $\kappa$ and not \ac{RDF}, and which uses local variable names
and not \acp{URI}, so the materialised facts are transformed into a suitable
internal representation. Substitution into templates is done, and finally some
post-processing is done to derive any remaining program directive that require
the complete assembled circuit in order to know.

It is interesting to consider that the entire compiler can be thought of as
implementing a kind of inference quite different from what is commonly used with
the Semantic Web. The consequent, the executable model, is in a different
language from the antecedent, the declarative description. Through the use of
embedding annotations, however, the original model is nevertheless carried
through to the output, and is unambiguously recoverable. There is thus an arrow
from the space of declarative models in RDF to the space of annotated executable
models. There is an arrow in the other direction that forgets the executable
part and retains the declarative part. In an important sense, the two
representations contain the same information, only that the executable model has
more materialised detail in order that it may be run.

\subsection{Semantic Inference}
\label{sec:inference-rules}

The input from the user is the model description in the high-level language as
described in Section~\ref{sec:gcdl}. This description uses terms from, and
makes reference to the \term{gcc} and \term{rbmo} vocabularies. The
\emph{meaning} of these terms, in the context of deriving an equivalent version
of the program in the low-level language, is given by the companion inference
rules. This is a somewhat subtle concept so let us illustrate what it means.
Consider the statement,
\begin{lstkappa}
  :R0040a a rbmc:Operator.
\end{lstkappa}
This statement gives the type of \term{:R0040a} as \term{rbmc:Operator}. 

The implications of this statement allows to identify the correct template to
use for this part, found from information provided by the \term{gcc} vocabulary.
Indeed, as a background fact, we are told that
\begin{lstkappa}
  rbmc:Operator rbmc:kappaTemplate rbmt:operator.ka.
\end{lstkappa}
or in other words that an \term{rbmc:Operator} corresponds to the template
\term{rbmt:operator.ka}. We also have an inference rule, provided with the
\term{gcc} vocabulary that says,
\begin{lstkappa}
  { ?part a [ rbmc:kappaTemplate ?template ] } => { ?part rbmc:kappaTemplate ?template }.
\end{lstkappa}
In the Notation 3~\citep{berners-lee_n3logic:_2008,berners-lee_notation3_2011}
language this means that, ``for all
\term{?part}s that has a type that corresponds to a kappa
\term{?template}, that \term{?part} itself corresponds to that
\term{?template}''. Alternatively,
\begin{equation*}
  \mathtt{type}(p, x) \wedge \mathtt{kappa}(x, t)
  \rightarrow
  \mathtt{kappa}(p, t)
\end{equation*}

It would have been perfectly possible to explicitly write what template should
be used for each part in the high-level model description. It is not
desirable to do so because it leaks implementation details of the compiler into
what ought to be an implementation-independent declarative description.

The above rule, and others like it serve to elaborate the high-level description
into a more detailed version suitable for the next stage of the compiler. All
implications that can be drawn under the \term{rdfs} inference rules and the
\term{gcc} specific rules must be drawn and must become part of the in the
in-memory RDF storage as the transitive closure of the rules (given the
background facts and the provided model facts).

\subsection{Internal Representation}

The output of the first stage of the compiler contains all the information
necessary to completely describe the output, but it is not in a convenient form
for providing to the template rendering engine. In our case the implementation
language for the compiler is in Python and the chosen rendering engine is
Jinja2~\citep{ronacher_jinja2_2008}. This means that the  appropriate data-structure is a
dictionary or associative list that can be processed natively by these tools
without need of external library. The required internal representation is built
up by querying the in-memory RDF storage for the specific information required
by the templates.


Our implementation does not require modification when new terms are added to the vocabulary and templates.
Suppose that a new kind of part is invented. This should mean writing a new
template for it and possibly adding some terms to the vocabulary but should not
require changing the compiler software itself. What makes this possible are the
inference rules described in the previous Section~\ref{sec:inference-rules}. The queries
on the RDF storage that produce the internal representation are posed in
terms of the \emph{consequents} of the inference rules rather than the specific
form of input.

\subsection{Template Substitution}
\label{sec:templates}

The templates that produce the bulk of the low-level output are written in the
well-known Jinja2 language. This language is commonly used for the server-side
generation of web pages. KaSim or BNGL programs are not web pages but they are
text documents and Jinja2 is well suited to generating them. It has a notion of
inclusion and inheritance that is useful for handling the variations among the
different kinds of parts, which typically differ in the rules for one or two of
the interactions in which they participate with the others being identical.

A full description of the facilities provided by Jinja2 is beyond the scope of
this paper, but a flavour is given in Figure~\ref{fig:template-example} which
\begin{figure}[ht]
  \begin{lstlisting}
## Auto-generated generic part {{ name }}
{% include "header.ka" %}
{% import "context.ka" as context with context %}
{% import "meta.ka" as meta with context %}

{% include "transcription_elongation.ka" %}
{% include "transcription_termination.ka" %}
{% include "translation_chain.ka" %}
{% include "translation_elongation.ka" %}
{% include "translation_termination.ka" %}
{% include "host_maintenance.ka" %}    
\end{lstlisting}
\begin{lstlisting}
{% set rule = "%s-translation-chain" % name %}
#
#^ :{{ rule }} a rbmo:Rule;
#^   bqbiol:isVersionOf go:GO:0006415;
{{ meta.rule() }}{# #}
#^   rdfs:label "{{ name }} formation of \
#^               translational chains, due to \
#^               gene fusion or leakiness of \
#^               stop codons".
# {{ name }} formation of translational chains, 
# due to gene fusion or leakiness of stop codons
#
'{{ rule }}' \
    RNA(ds!2, bs!1), \
    Ribosome(rna!1, protein!3), \
    RNA(type~{{ name }}, us!2, bs), \
    P(ds, bs!3) \
-> \
    RNA(ds!2, bs), \
    Ribosome(rna!1, protein!3), \
    RNA(type~{{ name }}, us!2, bs!1), \
    P(ds!4, bs), \
    P(type~{{ name }}, us!4, bs!3) \
    @{{ translationElongationRate }}
\end{lstlisting}
  \caption{Template examples. On top is the template for a generic part, and it
    references several other templates, one of which,
    \lstinline{translation_chain.ka}, is reproduced on bottom.}
     \label{fig:template-example}
\end{figure}
shows an example of a template for a generic part (not having specific
functionality like a promoter or operator might) demonstrating substitution
of the \lstinline{name} variable derived from annotation, and include
statements referencing several other templates, one of which is reproduced
and shows the actual KaSim code that is produced.

We use specific terms for defining the rates for the rules in which biological
parts are involved, and a few other terms according to the function of the
biological part of interest. It is possible to find the available terms out by
inspecting the \term{gcc} vocabulary.

\begin{figure}
  \begin{lstturtle}
  rbmc:transcriptionFactor a rbmc:Token;
      skos:prefLabel "transcriptionFactor".
  rbmc:transcriptionFactorBindingRate a rbmc:Token;
      skos:prefLabel "transcriptionFactorBindingRate";
      rbmc:default 1.0.
  rbmc:transcriptionFactorUnbindingRate a rbmc:Token;
      skos:prefLabel "transcriptionFactorUnbindingRate";
      rbmc:default 1.0.

  rbmc:Operator rdfs:subClassOf rbmc:Part;
      rbmc:kappaTemplate rbmt:operator.ka;
      rbmc:bnglTemplate rbmt:operator.bngl;
      rbmc:tokens
          rbmc:transcriptionFactor,
          rbmc:transcriptionFactorBindingRate,
          rbmc:transcriptionFactorUnbindingRate.
  \end{lstturtle}
  \caption{The specification in the \term{rbmc} vocabulary of an
    \term{rbmc:Operator} and associated terms.}
  \label{fig:gcc-part}
\end{figure}
A fragment of the \term{gcc} vocabulary is reproduced in
Figure~\ref{fig:gcc-part}. Though this exposes some implementation detail, it
is useful to understand the relationships between the various terms used to
describe models. This is also important when supplying customised templates.

There are \term{gcc:Tokens}, so named because they correspond to tokens
in the low-level language that are replaced. Each must have a preferred label that gives the
literal token. In cases where there exists a sensible default value, this
is given with \term{rbmc:default}. The purpose of these statements is to act as
a bridge between the fully materialised \ac{RDF} representation of the model
and the templates that require substitution of locally meaningful names.

For each kind of part (such as the \term{rbmc:Operator} in the example in
Figure~\ref{fig:gcc-part}), 
there are two main annotations that are necessary. For each machine-readable
low-level language, a template is specified. The \term{gcc:tokens}
annotations give the tokens that are pertinent to this kind of part. These must
be specified in the high-level model or allowed to take on their default values.
In addition to documenting the requirements of the templates for each kind of
part, these statements are, ``operationalised'' and used by the compiler. They
can equally well be used to check that a supplied high-level model is
well-formed.

\subsection{Derivation of Declarations}

The KaSim language requires forward declaration of the type signatures of agents.
This is by design~\citep{feret_personal_2015} so that the simulator can check
that agents are correctly used where they appear in patterns in the rules. While
this design choice can help a modeller that is writing a simulation program in
the low-level language by hand, to assist in finding mistakes and typographical
errors, it is not possible to know \emph{a priori} what these declarations
should be in the present context. The correct declarations for \texttt{DNA},
\texttt{RNA} and \texttt{Protein} depend on the complete set of parts that make
up the model so their correct declarations cannot be known in any template for
an individual part.

To solve this issue, the compiler implements a post-processing step. The rules that are
produced by instantiating the templates for each part are concatenated together
with any explicitly written rules that are to be included and then the whole is
parsed. The use of each agent in each rule in this rule-set is assumed to be
correct by definition. From there a declaration that covers each use of each
agent is built up.

\subsection{Initialisation}

At this final stage of the compiler, all rules are present, both supplied by the user for the host environment and implied by the
parts that form the genetic circuits and all declarations are also present. What
is missing is the statement that creates an initial copy of the DNA sequence
itself, which each upstream--downstream bond present. This information is, of
course, available in the definition of the circuit, and so an appropriate
\texttt{\%init} statement, creating an instance of the DNA sequence with correct
linkages between the agent-parts is produced and added to the output. The low-level
program is finally complete and ready to be executed.

\section{Future Work: Generating Models}
\label{sec:ga}

So far, we have demonstrated a computational pathway from a succinct model of a
genetic circuit written in RDF to an executable program for simulation or a
description in a language suited to consumption by laboratory tools. Recall that
the rationale for such an infrastructure is to bring computational tools to bear
on the design problem of genetic circuits: to identify circuits that are
feasible for some purpose through simulation in the first instance, before
attempting synthesis in the laboratory, for reasons of cost and accessibility.
The question which remains is, how do we know which circuits to model in the
first place?

A high-level design problem in genome synthesis is to create a circuit that
will respond in a particular way to a given chemical input. The desired
specification is given in terms of pairs of input and output measurements ---
typically time-series of concentrations of chemicals --- and
the circuit that implements the circuit is thought of as implementing this
mapping from input to output. 

One approach which does not use computation is to simply produce, in the
laboratory, all possible circuits from a given library of
parts~\citep{guet2002combinatorial,menzella2005combinatorial,smanski2014functional,cress2015crispathbrick}.
In that strategy, DNA sequences are randomly assembled into all possible
combinations and one simply looks for output markers such as the commonly used
fluorescent green protein that appear or not in response to given input. When a
cell that has the desired behaviour is found, its genome is sequenced to find
out what combination of parts it has implemented. This approach has the great
advantage of massive parallelism: all possibilities can be attempted in parallel
and only those candidate genomes that appear to be of interest analysed further.
It has the disadvantage of requiring expensive, specialised equipment to
identify potentially feasible solutions.

To bring the design problem into the computational realm, we propose the use of
an evolutionary algorithm. The process, which we only briefly sketch here, is as
follows. We begin with a specification, a set of (input, output) pairs that
describe the desired function of the system. We require a metric suitable for
measuring the distance between two such sets. This allows evaluation of the
feasibility of a circuit under test. With the problem so specified, we may begin
the search for suitable candidates.

To generate circuits, we use a library of parts from a database that have their
rates and other characteristics specified as outlined above. A large class of
circuits can be immediately eliminated --- those that are ill-formed. We use a
Pedersen-style grammar~\citep{pedersen_towards_2009} to only generate
well-formed circuits. We begin with an arbitrary circuit, perhaps one that is
thought to implement a function that is in some way similar to the desired one,
or perhaps one that is completely arbitrary. The circuit is then simulated
enough times to obtain a distribution of (input, output) pairs --- recall that
the model is necessarily stochastic. Now substitute one part in the circuit with
a different part of the same type from the library and simulate that. If the
(empirically) expected result is closer to the specification, adopt the new
circuit as the best candidate. Continue substituting parts until no further
improvement is found. Occasionally make larger changes, substituting more than
one part, in order that this optimisation process does not become trapped in
a local minimum.

This strategy for identifying feasible genetic circuits is, of course, a
\emph{genetic algorithm}. It imitates the evolutionary process of living
organisms in order to find the optimal solution to a given problem. Despite it
being far more efficient than an exhaustive search of the space of possible
circuits, it may yet be too computationally expensive. More ways to constrain
the search space will need to be developed. Nevertheless there is a certain
elegance to using a nature-inspired computational technique to design genomes
\emph{in silico} for later construction \emph{in vitro}.

\section{Conclusion}

We have, in this paper, presented the \acs{GCDL}, a high-level Semantic Web
language for describing genetic circuits. We have shown how these circuits
can be implemented in a rule-based language, and described our compiler that
translates between the description and the low-level implementation is
implemented taking full advantage of inference rules to maintain succinctness.
Finally we have sketched an area of future research in potential solution to the
problem of discovering such circuits in the first place.

\paragraph{Acknowledgements}
Thanks to Michael Korbakov for porting the original Haskell implementation
of the agent declaration code to Python.

\paragraph{Funding}
W.W., M.C., G.M., A.W., and V.D. acknowledge the support from the Engineering
and Physical Sciences Research Council (EPSRC) grant EP/J02175X/1 and from UK
Research Councils’ Synthetic Biology for Growth programme.
W.W. and V.D. acknowledge the European Union's Seventh Framework Programme for
research, technological development and demonstration grant number 320823 (to
V.D. and W.W.).
W.W. also acknowledges support from the National Academies Keck Futures
Initiative of the National Academy of Sciences award number NAKFI CB12.

\paragraph{Conflicts of Interest} The authors have no competing interests.

\bibliographystyle{apalike}
\bibliography{Compiler}
\appendix
\clearpage
\section{The Genetic Circuit Compiler Language}
\label{sec:gcc-vocabulary}
\begin{lstlisting}[language=turtle]
# -*- n3 -*-
@prefix dct:  <http://purl.org/dc/terms/>.
@prefix foaf: <http://xmlns.com/foaf/0.1/>.
@prefix owl:  <http://www.w3.org/2002/07/owl#> .
@prefix prov: <http://www.w3.org/ns/prov#>.
@prefix rbmo: <http://purl.org/rbm/rbmo#>.
@prefix gcc: <http://purl.org/rbm/comp#>.
@prefix rbmt: <http://purl.org/rbm/templates/>.
@prefix rdfs: <http://www.w3.org/2000/01/rdf-schema#>.
@prefix skos: <http://www.w3.org/2004/02/skos/core#>.

gcc:part a gcc:Token; skos:prefLabel "name".
gcc:Part a owl:Class;
    gcc:tokens gcc:part.

gcc:next a gcc:Token; skos:prefLabel "next".

gcc:transcriptionFactor a gcc:Token;
    skos:prefLabel "transcriptionFactor";
    gcc:default 1.0.
gcc:transcriptionFactorBindingRate a gcc:Token;
    skos:prefLabel "transcriptionFactorBindingRate";
    gcc:default 1.0.
gcc:transcriptionFactorUnbindingRate a gcc:Token;
    skos:prefLabel "transcriptionFactorUnbindingRate";
    gcc:default 1.0.

gcc:rnapBindingRate a gcc:Token;
    skos:prefLabel "rnapBindingRate";
    gcc:default 1.0.
gcc:rnapDNAUnbindingRate a gcc:Token;
    skos:prefLabel "rnapDNAUnbindingRate";
    gcc:default 1.0.
gcc:rnapRNAUnbindingRate a gcc:Token;
    skos:prefLabel "rnapRNAUnbindingRate";
    gcc:default 1.0.

gcc:ribosomeBindingRate a gcc:Token;
    skos:prefLabel "ribosomeBindingRate";
    gcc:default 1.0.
gcc:ribosomeRNAUnbindingRate a gcc:Token;
    skos:prefLabel "ribosomeRNAUnbindingRate";
    gcc:default 1.0.
gcc:ribosomeProteinUnbindingRate a gcc:Token;
    skos:prefLabel "ribosomeProteinUnbindingRate";
    gcc:default 1.0.

gcc:transcriptionInitiationRate a gcc:Token;
    skos:prefLabel "transcriptionInitiationRate";
    gcc:default 1.0.
gcc:transcriptionElongationRate a gcc:Token;
    skos:prefLabel "transcriptionElongationRate";
    gcc:default 1.0.

gcc:translationElongationRate a gcc:Token;
    skos:prefLabel "translationElongationRate";
    gcc:default 1.0.

gcc:rnaDegradationRate a gcc:Token;
    skos:prefLabel "rnaDegradationRate";
    gcc:default 1.0.
gcc:proteinDegradationRate a gcc:Token;
    skos:prefLabel "proteinDegradationRate";
    gcc:default 1.0.

gcc:Operator rdfs:subClassOf gcc:Part;
    gcc:kappaTemplate rbmt:operator.ka;
    gcc:bnglTemplate rbmt:operator.bngl;
    gcc:tokens
        gcc:transcriptionFactor,
        gcc:transcriptionFactorBindingRate,
        gcc:transcriptionFactorUnbindingRate,
        gcc:rnapDNAUnbindingRate,
        gcc:rnapRNAUnbindingRate,
        gcc:transcriptionInitiationRate,
        gcc:transcriptionElongationRate,
        gcc:ribosomeRNAUnbindingRate,
        gcc:ribosomeProteinUnbindingRate,
        gcc:translationElongationRate,
        gcc:rnaDegradationRate,
        gcc:proteinDegradationRate.

gcc:Promoter rdfs:subClassOf gcc:Part;
    gcc:kappaTemplate rbmt:promoter.ka;
    gcc:bnglTemplate rbmt:promoter.bngl;
    gcc:tokens
        gcc:next,
        gcc:rnapBindingRate,
        gcc:rnapDNAUnbindingRate,
        gcc:rnapRNAUnbindingRate,
        gcc:transcriptionInitiationRate,
        gcc:transcriptionElongationRate,
        gcc:ribosomeRNAUnbindingRate,
        gcc:ribosomeProteinUnbindingRate,
        gcc:translationElongationRate,
        gcc:rnaDegradationRate,
        gcc:proteinDegradationRate.

gcc:RibosomeBindingSite rdfs:subClassOf gcc:Part;
    gcc:kappaTemplate rbmt:rbs.ka;
    gcc:bnglTemplate rbmt:rbs.bngl;
    gcc:tokens
        gcc:rnapDNAUnbindingRate,
        gcc:rnapRNAUnbindingRate,
        gcc:transcriptionElongationRate,
        gcc:ribosomeBindingRate,
        gcc:ribosomeRNAUnbindingRate,
        gcc:ribosomeProteinUnbindingRate,
        gcc:translationElongationRate,
        gcc:rnaDegradationRate,
        gcc:proteinDegradationRate.

gcc:protein a gcc:Token;
    skos:prefLabel "protein".

gcc:CodingSequence rdfs:subClassOf gcc:Part;
    gcc:kappaTemplate rbmt:cds.ka;
    gcc:bnglTemplate rbmt:cds.bngl;
    gcc:tokens
        gcc:protein,
        gcc:rnapDNAUnbindingRate,
        gcc:rnapRNAUnbindingRate,
        gcc:transcriptionElongationRate,
        gcc:ribosomeRNAUnbindingRate,
        gcc:ribosomeProteinUnbindingRate,
        gcc:translationElongationRate,
        gcc:rnaDegradationRate,
        gcc:proteinDegradationRate.

gcc:Terminator rdfs:subClassOf gcc:Part;
    gcc:kappaTemplate rbmt:generic.ka;
    gcc:bnglTemplate rbmt:generic.bngl;
    gcc:tokens
        gcc:rnapDNAUnbindingRate,
        gcc:rnapRNAUnbindingRate,
        gcc:transcriptionElongationRate,
        gcc:ribosomeRNAUnbindingRate,
        gcc:ribosomeProteinUnbindingRate,
        gcc:translationElongationRate,
        gcc:rnaDegradationRate,
        gcc:proteinDegradationRate.
\end{lstlisting}
\clearpage
\section{Additional Inference Rules for GCC}
\label{sec:gcc-rules}
\begin{lstlisting}[language=turtle]
# -*- n3 -*-
@prefix dct:  <http://purl.org/dc/terms/>.
@prefix foaf: <http://xmlns.com/foaf/0.1/>.
@prefix owl: <http://www.w3.org/2002/07/owl#> .
@prefix prov: <http://www.w3.org/ns/prov#>.
@prefix rbmo: <http://purl.org/rbm/rbmo#>.
@prefix gcc: <http://purl.org/rbm/comp#>.
@prefix rbmt: <http://purl.org/rbm/templates/>.
@prefix rdfs: <http://www.w3.org/2000/01/rdf-schema#>.
@prefix skos: <http://www.w3.org/2004/02/skos/core#>.

## The preferred label of a part is it's part slug
{ ?part gcc:part ?label } => { ?part skos:prefLabel ?label }.

## Derivation of templates
{ ?part a [ gcc:kappaTemplate ?template ] } => { ?part gcc:kappaTemplate ?template }.
{ ?part a [ gcc:bnglTemplate ?template ] } => { ?part gcc:bnglTemplate ?template }.

## Translation of special predicates to replacement instructions
{ ?kind gcc:tokens ?token .
  ?token skos:prefLabel ?label .
  ?part a ?kind; ?token ?value } =>
{ ?part gcc:replace [ gcc:string ?label; gcc:value ?value ] }.
\end{lstlisting}

\end{document}